\documentclass{jps-cp}
\usepackage{txfonts} %Please comment out this line unless the txfonts package is availabe in your LaTeX system.

\usepackage{bm}
\usepackage{calligra}
\usepackage{amsmath,amssymb}
\usepackage{braket}
\newcommand{\Jost}{\text{\textcalligra{f}\;}}
\newcommand{\JostM}{\text{\textcalligra{F}\;}}
\def\newblock{\hskip .11em plus .33em minus .07em}
\usepackage{color}

\title{Decay Effect on Near-Threshold Mass Scaling with Complex and Coupled-Channel Potentials}

\author{Erick \textsc{Gushiken}$^{1}$ and Tetsuo \textsc{Hyodo}$^{2}$}

\inst{$^{1}$Department of Physics, Tokyo Metropolitan University, Hachioji 192-0397, Japan \\
$^{2}$Research Center for Nuclear Physics, The University of Osaka, Ibaraki, Osaka 567-0047, Japan}

\email{hyodo@rcnp.osaka-u.ac.jp}

\abst{
We investigate the effect of decay channels on the near-threshold mass scaling by employing potential models. By varying the attractive strength of a square-well potential, we examine the pole trajectory associated with the transition of an $s$-wave bound state into a resonance state, incorporating decay-channel effects through both a single-channel complex potential model and a coupled-channel real potential model. As a result, we show that the pole of a quasibound state below the threshold is not continuously connected to that of a resonance state above the threshold. Furthermore, by comparing the results obtained from the single-channel and coupled-channel models, we clarify the correspondence between the pole trajectories in the two approaches.
}

\kword{Exotic hadrons, Coupled-channel scattering, Pole trajectories}

\begin{document}
\maketitle

%%%%%%%%%%%%%%%%%%%%
\section{Introduction}
\label{sec:intro}
%%%%%%%%%%%%%%%%%%%%

% introduction
Recent advances in experimental techniques have led to the observation of many exotic hadron candidates~\cite{ParticleDataGroup:2024cfk}. In response, increasing attention has been devoted to understanding the internal structure of hadrons, and extensive theoretical efforts have been made to explore new pictures beyond the conventional quark models~\cite{Guo:2017jvc,Brambilla:2019esw}. The existence of such hadrons beyond the traditional framework suggests that the strong interaction can give rise to various possible forms of matter.

% unstable hadrons
Most hadrons are unstable via the strong interaction~\cite{ParticleDataGroup:2024cfk}, as they decay through coupling to hadronic scattering channels with lower energies. The wave functions of such unstable states do not converge at large distances and therefore exhibit qualitatively different properties from those of stable bound states~\cite{Moiseyev,Hyodo:2020czb}. Consequently, to understand the internal structure of exotic hadrons, it is essential to take into account their instability by incorporating channel coupling in theoretical studies.

% bound to resonance transition
Exotic hadrons are often observed near two-hadron thresholds, where weakly bound states are known to be dominated by hadronic molecular components in their internal structure~\cite{Weinberg:1965zz,Hyodo:2014bda,Kamiya:2015aea,Hanhart:2022qxq,Kinugawa:2023fbf}. This feature is closely related to the near-threshold mass scaling that is observed when an $s$-wave bound state evolves into a resonance state through a virtual state~\cite{Ikeda:2011dx,Hyodo:2013iga,Hyodo:2014bda,Hanhart:2014ssa}. Corresponding trajectories of the poles of the scattering amplitude have also been observed in studies of the quark-mass dependence of hadron resonances~\cite{Hanhart:2008mx} and the partial restoration of chiral symmetry~\cite{Hyodo:2010jp}.

% this work
In this study, we investigate the effect of decay channels on this near-threshold mass scaling. It has been established that when a bound state is coupled to a decay scattering channel with lower energy, a quasibound state with a finite decay width, called the Feshbach resonance~\cite{Feshbach:1958nx,Feshbach:1962ut}, is formed. Recent analyses based on chiral effective models have discussed the influence of decay channels on the trajectories of the poles~\cite{Nishibuchi:2025uvt}. In the present work, we aim to clarify the effect of decay channels on the threshold mass scaling by employing coordinate space potential models. 

%%%%%%%%%%%%%%%%%%%%
\section{Formulation}
\label{sec:formulation}
%%%%%%%%%%%%%%%%%%%%

We adopt two approaches to incorporate the effects of decay channels into a potential model. One is the optical potential method, in which the effects of channel coupling are taken into account implicitly by introducing an imaginary part into the potential. The other approach explicitly incorporates the coupling to decay channels by solving the coupled-channel Schr\"odinger equation. 

%==========================
\subsection{Single-channel model}
\label{subsec:singleformulation}

% Schroedinger equation
Here we consider a two-body scattering system with the reduced mass $\mu$, interacting via a central force without explicit channel coupling. The $s$-wave radial Schr\"odinger equation is given by
\begin{align}
\left[-\frac{1}{2\mu}\frac{d^{2}}{dr^{2}}+(V_{0}+iW_{0})\Theta(r-b)\right]u(r)
& =
Eu(r) ,
\label{eq:Schroedinger}
\end{align}
where $u(r)$ is the radial wave function with energy $E$. We adopt a square-well potential of range $b$ with a complex strength $V_{0}+iW_{0}$. Here, $V_{0}$ represents the interaction strength, while $W_{0}$ accounts for absorption effects due to implicit decay channels. For scattering states with $E>0$, the momentum is defined as $p=\sqrt{2\mu E}$.

% discrete eigenstates
By solving the Schr\"odinger equation~\eqref{eq:Schroedinger} with an outgoing boundary condition, one obtains the discrete eigenstates of the Hamiltonian, which are equivalently identified by the zeros of the Jost function~\cite{Hyodo:2020czb,Taylor,Rakityansky}. The Jost function $\Jost(p)$ is defined as the coefficient of the incoming wave in the asymptotic form of the regular solution, which is the scattering solution satisfying the boundary condition $\lim_{r\to 0}[u(r)/\sin(pr)]=1$ at the origin. Therefore, by analytically continuing the Jost function into the complex momentum plane, the solutions of $\Jost(p)=0$ give the eigenmomenta of the discrete eigenstates. For the square-well potential in Eq.~\eqref{eq:Schroedinger}, the Jost function is given analytically by
\begin{align}
\Jost(p)
&=\left[\cos\left(\sqrt{p^{2}-2\mu(V_0+iW_{0})}, b\right)
-i\frac{p\sin\left(\sqrt{p^{2}-2\mu(V_0+iW_{0})}, b\right)}{\sqrt{p^{2}-2\mu(V_0+iW_{0})}}
\right]e^{ipb}.
\end{align}
Because zeros of the Jost function correspond to poles of the scattering amplitude and the $S$-matrix, solutions of $\Jost(p)=0$ are referred to as poles in what follows. In general, poles located on the positive imaginary axis of the complex momentum plane correspond to bound states, those on the negative imaginary axis to virtual states, and those in the fourth (third) quadrant to resonances (anti-resonances). For a real potential with $W_{0}=0$, the solutions appear only on the positive imaginary axis or in the lower half-plane, whereas for a complex potential with $W_{0}\neq 0$, the solutions of $\Jost(p)=0$ can exist over the entire complex $p$ plane.

%==========================
\subsection{Coupled-channel model}
\label{subsec:coupledformulation}

% discrete eigenstates
Next, we introduce a coupled-channel model to treat decay channels explicitly. For channels $i=1,2$, the $s$-wave radial Schr\"odinger equation is given by
\begin{align}
\left[
\left(-\frac{1}{2\mu}\frac{d^{2}}{dr^{2}}+\Delta_{i}
\right)\delta_{ij}
+V_{ij}\Theta(r-b)\right]u_{j}(r)
& =
Eu_{i}(r),
\label{eq:SchroedingerCC}
\end{align}
where $u_{i}(r)$ and $\Delta_{i}$ are the wave function and threshold energy of channel $i$, respectively. In this study, we focus on the near-threshold region of channel 2, and therefore choose the energy origin at this threshold, $\Delta_{1}<0$ and $\Delta_{2}=0$. For $E>0$, the momentum in each channel is defined as $p_{i}=\sqrt{2\mu(E-\Delta_{i})}$. The interaction is described by a square-well potential with real strengths $V_{ij}$. In the following, we assume that the decay channel is noninteracting ($V_{11}=0$) and that the system is invariant under time reversal, implying $V_{21}=V_{12}$.

% Effective single-channel potential
By applying the Feshbach projection method~\cite{Feshbach:1958nx,Feshbach:1962ut} to the coupled-channel Hamiltonian and eliminating the decay channel, we obtain a nonlocal and energy-dependent effective potential,
\begin{align}
   V_{\rm eff}^{(\pm)}(E;\bm{r},\bm{r}^{\prime})
   &=-\frac{\mu V_{12}^{2}}{2\pi}
   \frac{e^{-\sqrt{2\mu(E-\Delta_{1}\pm i0^{+})}|\bm{r}-\bm{r}^{\prime}|}}{|\bm{r}-\bm{r}^{\prime}|}
   \Theta(b-|\bm{r}|)\Theta(b-|\bm{r}^{\prime}|),
\end{align}
where the superscript $\pm$ specifies the boundary condition of the Green's function of channel~1. The local part of this potential, obtained in the limit $\bm{r}\to\bm{r}^{\prime}$, takes the form
\begin{align}
   \lim_{\bm{r}\to \bm{r}^{\prime}}V_{\rm eff}^{(\pm)}(E;\bm{r},\bm{r}^{\prime})
   &=\mp i\frac{\mu V_{12}^{2}}{2\pi}\sqrt{2\mu(E-\Delta_{1})}
   \Theta(b-|\bm{r}|)
   +\mathcal{O}(|\bm{r}-\bm{r}^{\prime}|)
   \quad (E>\Delta),
\end{align}
for energies above the threshold of channel~1, $E>\Delta$. Therefore, choosing the conventional boundary condition ($+i0^{+}$) leads to a negative imaginary part of the effective potential, while reversing the boundary condition results in a sign flip of the imaginary part. As discussed in Ref.~\cite{Kamiya:2022thy}, the sign of the imaginary part of a complex potential is associated with the boundary condition of the eliminated channel, and the above result provides an explicit realization of this correspondence.

% discrete eigenstates 
In the case of two-channel scattering, the coupled differential equations~\eqref{eq:SchroedingerCC} admit two linearly independent regular solutions. The Jost matrix $\JostM_{ij}(E)$ is defined from the coefficients of the incoming waves in the asymptotic form of a square matrix constructed by arranging these two regular solutions, each of which has two components~\cite{Taylor,Rakityansky}. The discrete eigenstates in coupled-channel scattering are then obtained from the condition that the determinant of the Jost matrix vanishes,
\begin{align}
\det \left[ \JostM(E)\right]
=0 .
\label{eq:polecondcoupled}
\end{align}
For the square-well potential in Eq.~\eqref{eq:SchroedingerCC}, although the explicit expressions are rather involved, the Jost matrix can be obtained analytically in terms of $V_{ij}$, $b$, $\mu$, and $\Delta_{i}$~\cite{Rakityansky}.

% Riemann sheets
Here we summarize the Riemann-sheet structure of the complex plane for two-channel scattering. In general, the energy is a double-valued function of the momentum, and therefore the Riemann-sheet structure of the energy plane must be taken into account when analytically continuing quantities such as the Jost function. For single-channel scattering, the upper half-plane of the momentum corresponds to the first Riemann sheet ([t] sheet), while the lower half-plane corresponds to the second Riemann sheet ([b] sheet). In the case of two-channel scattering, the relation between the energy and the momentum must be specified for each channel, and the Riemann sheets are accordingly labeled as [tt], [tb], [bt], and [bb] as in Refs.~\cite{Pearce:1988rk,Nishibuchi:2023acl}. Thus, when the energy $E$ is taken as the variable, four Riemann sheets must be considered in two-channel scattering. On the other hand, if the momentum $p_{2}$ of channel 2 is used as the variable instead of $E$, the choice of the Riemann sheet is required only for $p_{1}$, while no specification is needed for $p_{2}$. In this case, the entire structure can be described by two Riemann sheets~\cite{Nishibuchi:2023acl}. However, the complex $p_{2}$ plane classified solely by the Riemann sheet of channel 1 has a branch cut along the real axis, which is inconvenient for tracing pole trajectories. Therefore, in the following, we employ the [tt/bb] and [bt/tb] sheets of $p_{2}$, which are constructed by interchanging the lower half-planes of the two momentum planes~\cite{Nishibuchi:2023acl}.

%%%%%%%%%%%%%%%%%%%%
\section{Numerical results}
\label{sec:results}
%%%%%%%%%%%%%%%%%%%%

%==========================
\subsection{Single-channel model}
\label{subsec:singleresults}

% real potential
We first consider the single-channel model introduced in Sec.~\ref{subsec:singleformulation} with a real potential, i.e., $W_{0}=0$. In this case, the number of bound states depends on the magnitude of the attractive strength $V_{0}$, and the critical strength at which the $n$-th bound state appears is given by $V_{0}^{(n)}=-\pi^{2}(2n-1)^{2}/8\ [b^{-2}\mu^{-1}]$. Specifically, the first bound state emerges at $V_{0}^{(1)}\simeq -1.23\,[b^{-2}\mu^{-1}]$, while the second bound state appears at $V_{0}^{(2)}\simeq -11.1\,[b^{-2}\mu^{-1}]$. In this study, we aim to investigate the near-threshold mass scaling associated with the transition from a bound state to a resonance as the attractive interaction is weakened. However, the $n=1$ bound state does not evolve into a resonance, since it does not have a corresponding virtual state. Therefore, we focus on the pole behavior near $V_{0}^{(2)}\simeq -11.1\,[b^{-2}\mu^{-1}]$, where the $n=2$ bound state undergoes a transition into a resonance state.

% V0 region
For $V_{0}=-15\,[b^{-2}\mu^{-1}]$, the pole corresponding to the $n=2$ bound state is located at $p\sim 2i\,[b^{-1}]$, while its accompanying virtual-state pole appears at $p\sim -4i\,[b^{-1}]$. As the attractive interaction is weakened by reducing $|V_{0}|$, these two poles move toward each other. At $V_{0}=V_{0}^{(2)}$, the bound-state pole reaches the negative imaginary axis and turns into a virtual state. With a further reduction of $|V_{0}|$, the two poles collide on the negative imaginary axis and subsequently evolve into a resonance ($R$) and an anti-resonance ($\bar{R}$). This behavior corresponds to the conventional near-threshold mass scaling observed for $s$-wave states~\cite{Hyodo:2014bda,Hanhart:2014ssa}.

\begin{figure}[tb]
\centering
\includegraphics[width=0.45\textwidth]{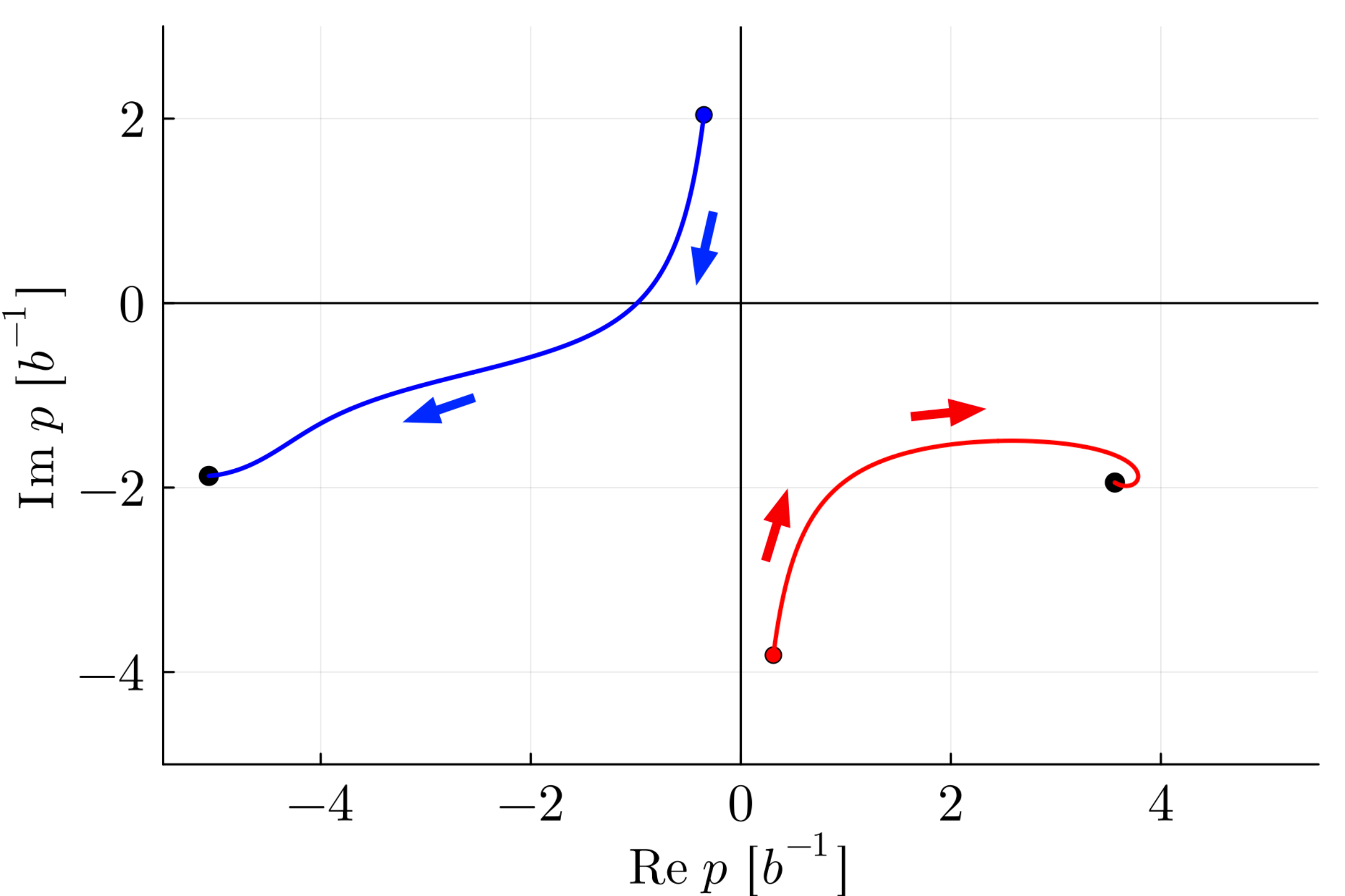}
\caption{Pole trajectories in the complex momentum plane by the single-channel model with $-15 \leq V_{0}\leq -0.1\ [b^{-2}\mu^{-1}]$ and $W_{0}=-1\ [b^{-2}\mu^{-1}]$.}
\label{fig:single}
\end{figure}

% introduction of W0
Figure~\ref{fig:single} shows the pole trajectories in the complex momentum plane when an imaginary part $W_{0}=-1\,[b^{-2}\mu^{-1}]$ is introduced in the potential. With the inclusion of the imaginary component, the bound-state pole moves into the region ${\rm Re}\,p<0$ and becomes a quasibound state ($QB$), while the virtual-state pole shifts into the region ${\rm Re}\,p>0$ and turns into a quasivirtual state ($QV$)~\cite{Nishibuchi:2023acl,Nishibuchi:2025uvt}. Starting from these poles, we reduce $|V_{0}|$ and find that the $QB$ pole evolves into a pole located in the third quadrant of the complex momentum plane. In contrast, the resonance pole is evolved from the $QV$ pole. These results indicate that, when the attractive interaction is varied and a quasibound state below the threshold evolves into a resonance state above the threshold, the poles representing these states are not continuously connected. Instead, a kind of interchange of the relevant poles takes place.

% non-Hermitian
We also note that the pole trajectories shown in Fig.~\ref{fig:single} are asymmetric with respect to the imaginary axis. This asymmetry originates from the non-Hermitian nature of the Hamiltonian due to the complex potential. As shown in Sec.~\ref{subsec:coupledformulation}, reversing the sign of the imaginary part in the complex effective potential leads to a solution with time-reversed boundary conditions in the decay channel. Indeed, when we perform the calculation with a positive imaginary part $W_{0}=+1\,[b^{-2}\mu^{-1}]$, we obtain pole trajectories that are reflected with respect to the imaginary axis compared to those in Fig.~\ref{fig:single}.

%==========================
\subsection{Coupled-channel model}
\label{subsec:coupledresults}

% V_{12} dependence
Next, we analyze the pole trajectories using the coupled-channel model introduced in Sec.~\ref{subsec:coupledformulation}. Focusing on the region near the threshold of channel~2, the interaction strength $V_{22}$ corresponds to $V_{0}$ in the single-channel model, while the transition interaction $V_{12}$ plays a role analogous to the decay parameter $W_{0}$. We choose the momentum $p_{2}$ of channel~2 as the complex variable, and consider the [tt/bb] and [bt/tb] sheets, as discussed above. However, in the absence of channel coupling ($V_{12}=0$), the choice of the Riemann sheet for channel~1 does not affect the result, and identical pole trajectories are obtained on both sheets. When a finite coupling $V_{12}$ is introduced, the bound state obtained at $V_{22}=-15\,[b^{-2}\mu^{-1}]$ turns into a quasibound state  ($QB$) and its conjugate ($\overline{QB}$) on the [bt/tb] sheet, while the virtual state becomes a quasivirtual state ($QV$) and its conjugate ($\overline{QV}$) on the [tt/bb] sheet. Since the Hamiltonian of the coupled-channel model is Hermitian, the poles in each momentum plane appear symmetrically with respect to the imaginary axis. With channel coupling, a branch cut appears on the imaginary axis in the region $-\sqrt{2\mu |\Delta|} < {\rm Im}\, p < \sqrt{2\mu |\Delta|}$. Therefore, the bound state on the [tt/bb] sheet can be interpreted as having crossed the cut and moved to the [bt/tb] sheet, while the virtual state on the [bt/tb] sheet moves to the [tt/bb] sheet~\cite{Nishibuchi:2025uvt}.

\begin{figure}[tb]
\centering
\includegraphics[width=0.45\textwidth]{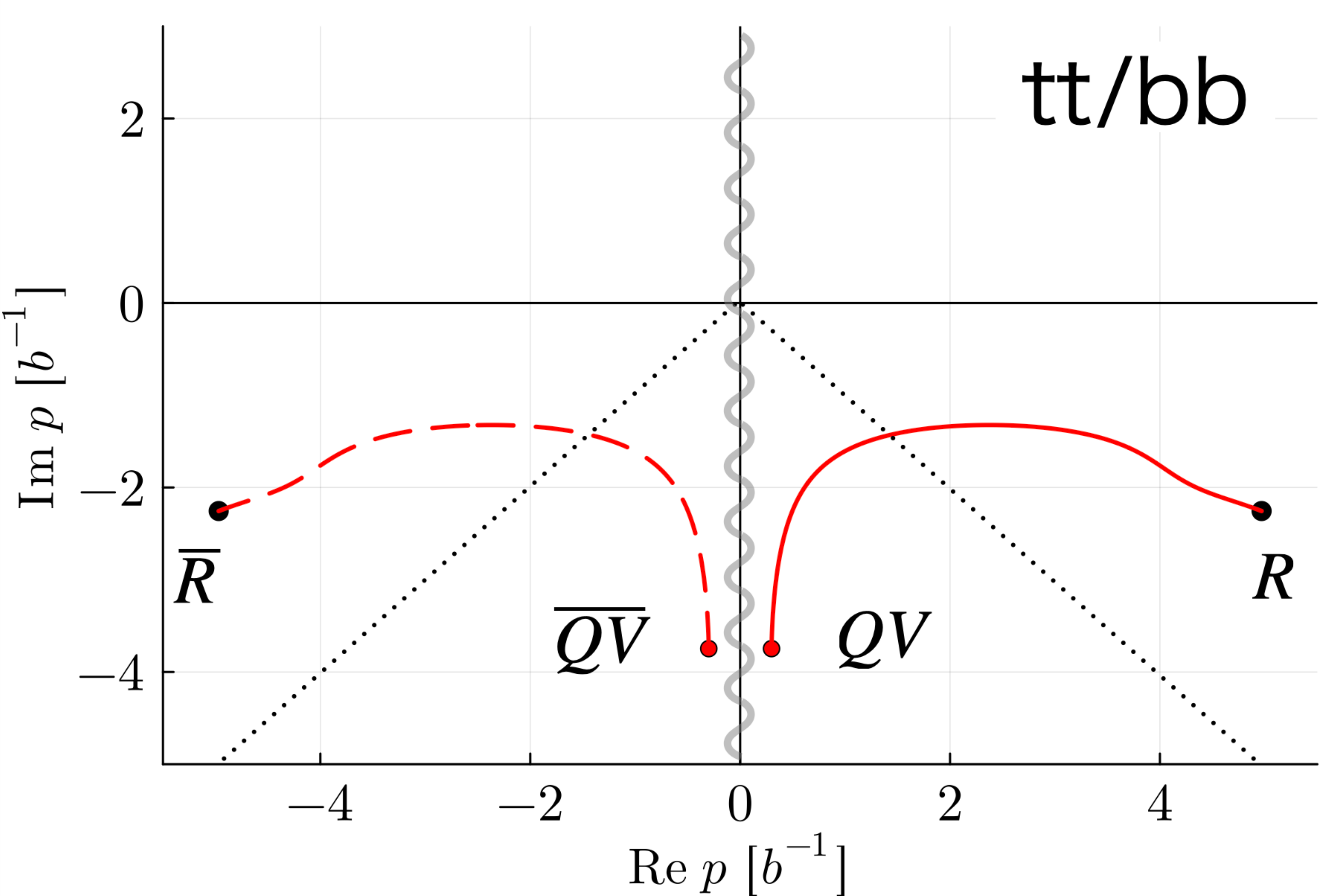}
\includegraphics[width=0.45\textwidth]{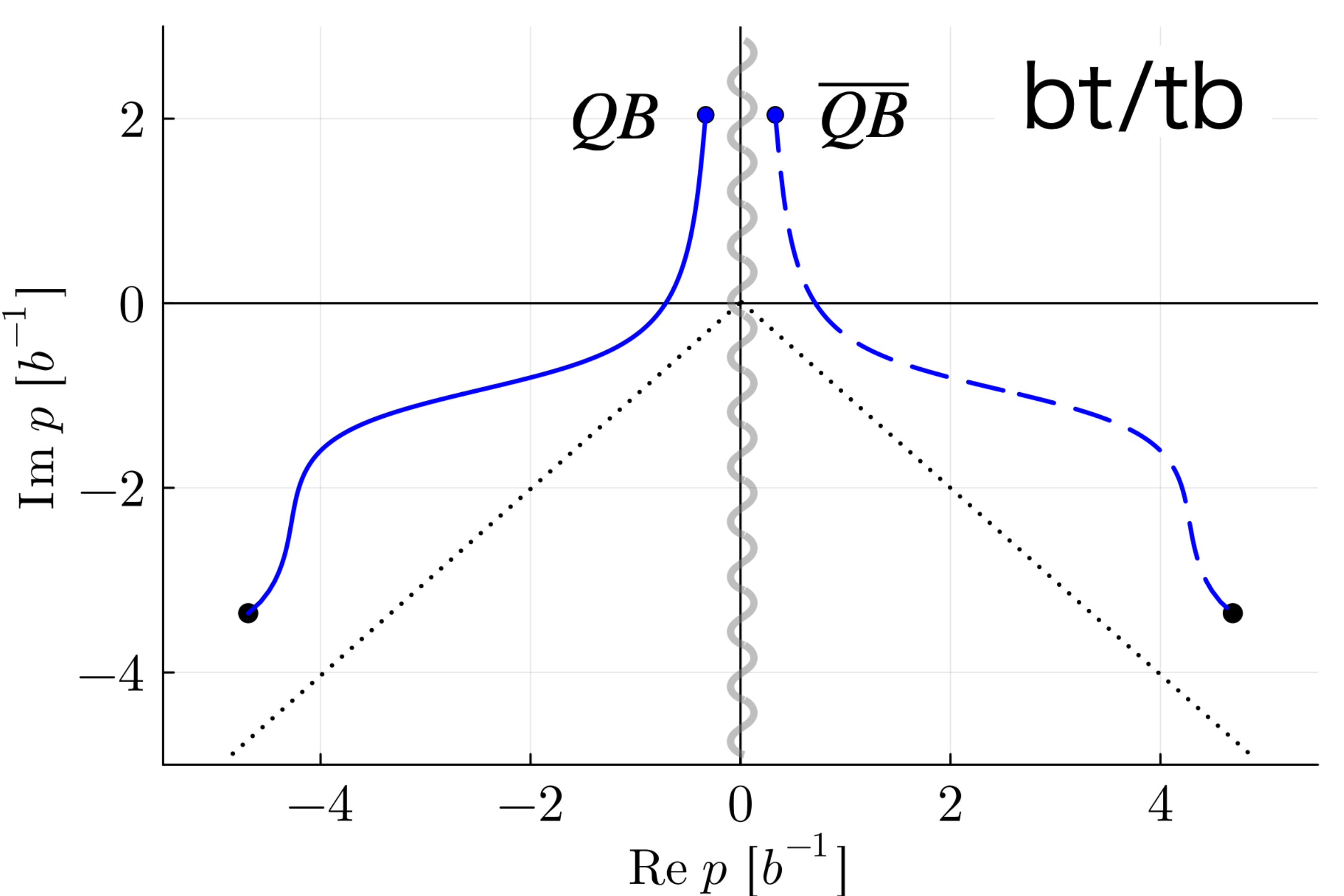}
\caption{Pole trajectories in the complex momentum plane by the coupled-channel model with  $-15 \leq V_{0}\leq -0.1\ [b^{-2}\mu^{-1}]$ and $V_{12}=2\ [b^{-2}\mu^{-1}]$. Left: [tt/bb] sheet, right: [bt/tb] sheet.}
\label{fig:coupled}
\end{figure}

% V_{22} dependence
For $V_{12}=2\ [b^{-2}\mu^{-1}]$ and $\Delta_{1}=-15\ [b^{-2}\mu^{-1}]$, the pole trajectories obtained by varying $|V_{22}|$ are plotted in Fig.~\ref{fig:coupled}. The pole of the quasibound state on the [bt/tb] sheet moves into the third quadrant of the complex momentum plane, while the pole of the quasivirtual state on the [tt/bb] sheet evolves into a resonance ($R$). Their respective conjugate poles trace trajectories that are symmetric with respect to the imaginary axis. As in the single-channel calculation shown in Fig.~\ref{fig:single}, the origin of the resonance pole can be traced back to the quasivirtual state located on the [bb] sheet, and it is not continuously connected to the quasibound state on the [bt] sheet. Similar pole trajectories have also been obtained in analyses based on chiral effective models \cite{Nishibuchi:2025uvt}.

\begin{figure}[tb]
\centering
\includegraphics[width=0.45\textwidth]{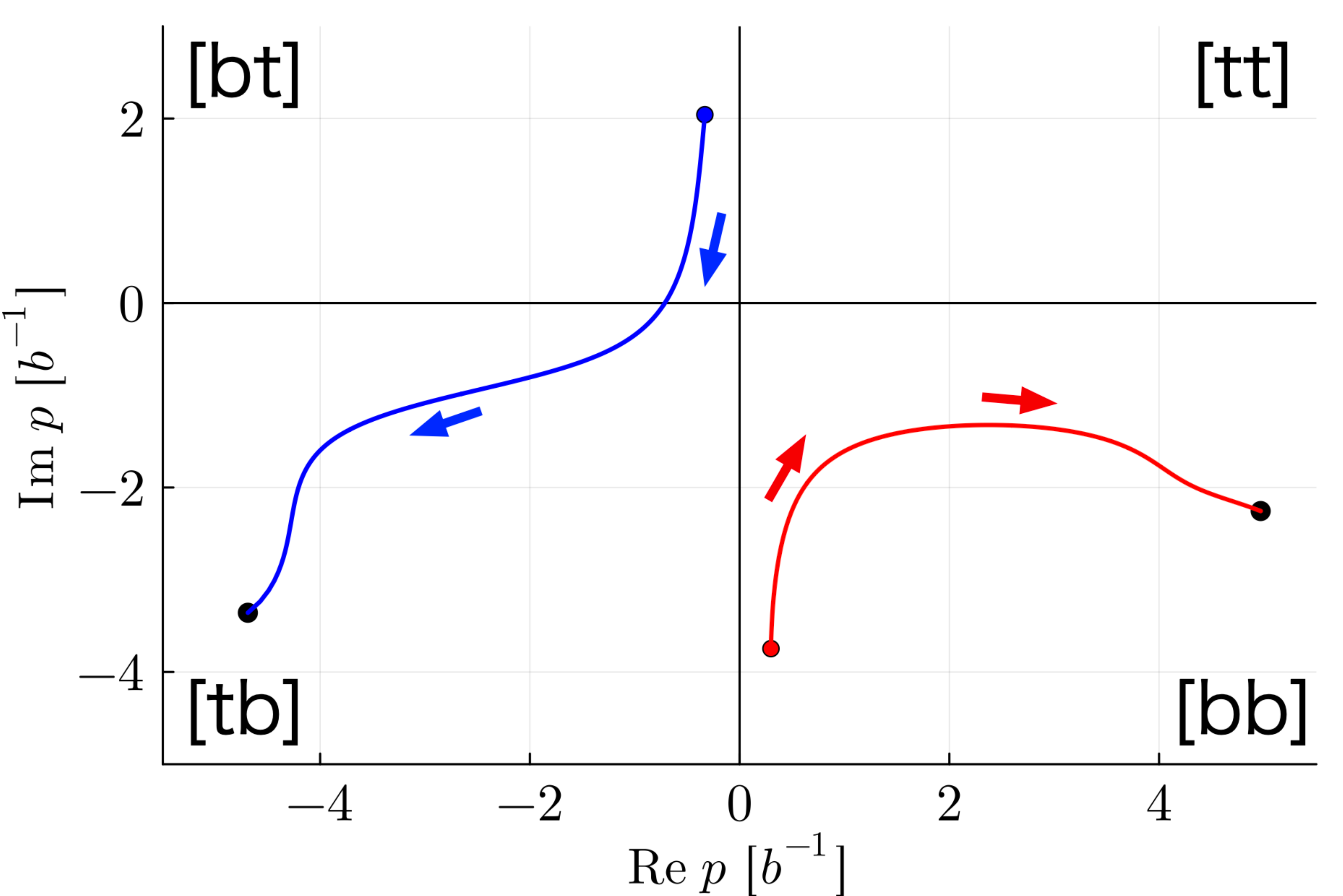}
\caption{Pole trajectories in the complex momentum plane by the coupled-channel model with  $-15 \leq V_{0}\leq -0.1\ [b^{-2}\mu^{-1}]$ and $V_{12}=2\ [b^{-2}\mu^{-1}]$. Left (right) half plane is the [bt/tb] ([tt/bb]) sheet.}
\label{fig:comparison}
\end{figure}

% Comparison of two approaches
By focusing on the trajectories of the $QB$ and $QV$ poles in Fig.~\ref{fig:coupled}, one finds that they exhibit qualitatively the same behavior as those shown in Fig.~\ref{fig:single}. Indeed, if the Riemann sheets are arranged such that the left half-plane corresponds to the [bt/tb] sheet and the right half-plane to the [tt/bb] sheet (Fig.~\ref{fig:comparison}), the resulting pole trajectories are seen to correspond directly to those obtained by the single-channel model. With this arrangement of the Riemann sheets, the cuts originating from the branch points $p=\pm i\sqrt{2\mu |\Delta|}$ extend toward $\pm i\infty$, and no cut appears in the vicinity of the origin $p=0$. In other words, the arrangement of the Riemann sheets shown in Fig.~\ref{fig:comparison} corresponds to selecting the Riemann sheets that are closest to the physical scattering region defined by $p_{2}>0$. Moreover, the pole trajectories that are symmetric with respect to the imaginary axis and obtained for $W_{0}>0$ can be interpreted, following the discussion in Ref.~\cite{Kamiya:2022thy}, as the time-reversed counterparts of the solutions shown in Fig.~\ref{fig:single}. In other words, by treating the decay channel explicitly within the coupled-channel model, the Hamiltonian remains Hermitian, allowing one to obtain not only the solutions that strongly affect physical scattering (the solid trajectories in Fig.~\ref{fig:coupled}), but also their time-reversal-symmetric counterparts (the dashed trajectories in Fig.~\ref{fig:coupled}) simultaneously.

%%%%%%%%%%%%%%%%%%%%
\section{Summary}
\label{sec:summary}
%%%%%%%%%%%%%%%%%%%%

In this study, we have investigated the trajectories of poles representing near-threshold discrete eigenstates in the complex momentum plane using square-well potential models. Focusing on the mass scaling in which a bound state turns into a resonance as the attractive interaction is weakened, we study the pole trajectories by incorporating the effects of decay channels. We find that the pole corresponding to a quasibound state is not continuously connected to the pole of a resonance. Furthermore, by comparing the results obtained from a single-channel complex potential model with those from a two-channel real potential model, we clarify that the choice of the sign of the imaginary part of the complex potential corresponds to the boundary conditions imposed on the eliminated decay channel. When the Riemann sheet that is most relevant to physical scattering is selected in the two-channel problem, pole trajectories qualitatively identical to those obtained
with the complex potential model are reproduced.

%
%\begin{table}[tbh]
%\caption{Captions to tables and figures should be sentences.}
%\label{t1}
%\begin{tabular}{ll}
%\hline
%AAA & BBB \\
%CCC & DDD \\
%\hline
%\end{tabular}
%\end{table}

%\begin{figure}[tbh]
%\includegraphics{fig01.eps}
%\caption{You can embed figures using the \texttt{\textbackslash includegraphics} command. Basically, figures should appear where they are cited in the text. You do not need to separate figures from the main text when you use \LaTeX\ for preparing your manuscript.}
%\label{f1}
%\end{figure}

%\bibliographystyle{h-physrev3}
%\bibliography{refs.bib}

%\begin{thebibliography}{9}
%\bibitem{cp} The abbreviation for JPS Conference Proceedings should be ``JPS Conf. Proc." in the reference list.
%\bibitem{jpsj} The abbreviation for the Journal of the Physical Society of Japan should be ``J. Phys. Soc. Jpn." in the reference list.
%\bibitem{ptep} The abbreviation for the Progress of Theoretical and Experimental Physics should be ``Prog. Theor. Exp. Phys." in the reference list.
%\bibitem{instructions} More abbreviations of journal titles are listed in ``Instructions for Preparation of Manuscript", which is available at our Web site (http://jpsj.jps.or.jp).
%\bibitem{format} F. Author, S. Author, and T. Author, Abbreviated journal title \textbf{volume in bold face}, initial page or article number (year of publication).
%\end{thebibliography}

\end{document}